\documentclass[prd,aps,twocolumn,nofootinbib]{revtex4}
\usepackage{graphicx}
\usepackage{xspace}
\usepackage{amsmath}
\def\nus{NuSTAR\xspace}

\usepackage{amsmath}

\begin{document}

\title{Decaying dark matter search with NuSTAR deep sky observations}
\author{A.~Neronov$^1$}
\author{D.~Malyshev$^2$}
\author{D.~Eckert$^1$}
\address{1. ISDC, Astronomy Department, University of Geneva, Ch. d'Ecogia 16, 1290, Version, Switzerland \\
2. Institut f{\"u}r Astronomie und Astrophysik T{\"u}bingen, Universit{\"a}t T{\"u}bingen, Sand 1, D-72076 T{\"u}bingen, Germany}
\begin{abstract}
We present the results of the search for decaying dark matter with particle mass in the  6-40~keV range with NuSTAR deep observations of COSMOS and ECDFS empty sky fields. We show that main contribution to the decaying dark matter signal from the Milky Way galaxy comes through the aperture of the NuSTAR detector, rather than through the focusing optics. High sensitivity of the NuSTAR detector, combined with the large aperture and large exposure times of the two observation fields allow us to improve previously existing constraints on the dark matter decay time by up to an order of magnitude in the mass range 10-30~keV. In the particular case of the $\nu$MSM sterile neutrino dark matter,  our constraints impose an upper bound $m<20$~keV on the dark matter particle mass.  We report detection of four unidentified spectral lines in our data set. These line detections are either due to the systematic effects (uncertainties of calibrations of the NuSTAR detectors) or have an astrophysical origin.  We discuss different possibilities for testing the nature of the detected lines.  \end{abstract}

\maketitle

\section{Introduction}
\label{sec:intro}

A range of particle models of the Dark Matter (DM) considers light weight (much below the proton and electron masses) DM particles which are unstable and could decay into the Standard Model particles with production of photons. The most known examples are sterile neutrinos (appearing e.g. in the $\nu$MSM model ~\cite{numsm1,numsm11,numsm,review,review1}) and axion-like particles \cite{axions}. The most clear observational signature of these models is the monoenergetic photon flux at the energy $E=m_{DM}/2$ ($m_{DM}$ is the DM particle mass) expected from all massive DM halos~\cite{pal,barger}. The strongest signal is generically expected to come from the Milky Way galaxy and is detectable from all the directions on the sky \cite{strategy}, with a moderate excess in the direction toward the inner galaxy, compared to the Galactic anti-centre direction. This means that all telescopes sensitive to photons with energies close to the DM particle mass could potentially be used as the DM detectors.  

If the DM particles are fermions, their mass is constrained to be heavier than $\sim 1$~keV, a constraint imposed by the phase space density of DM in the compact low mass DM halos~\cite{tremainegunn,dodelson,tg1,gorbunov08}. Somewhat tighter bounds arise from the non-observation of small scale structures suppression in Ly$\alpha$ forest data~\cite{lya1,lya2,lya3}. The decay signal from the fermionic DM is detectable with X-ray telescopes. A range of constraints on the lifetime of decaying DM (or on the mixing angle $\theta$ of sterile neutrino DM) have been previously derived from non-observation of the DM decay line by the X-ray telescopes \cite{constr1,constr2,constr3,constr4,constr5,constr6,constr7,spi,strategy,dima_thesis,riemer:14,horiuchi:14,dm_nustar_bullet,gbm,sekiya16}. An unidentified X-ray line at the energy 3.55~keV has been recently reported in the staked spectrum of galaxy clusters and in M31 galaxy~\cite{line35_1,line35_2}. Interpretation of this line as a DM decay line is in tension with the non-observation of the line in nearby dwarf spheroidal galaxies~\cite{dsphs,dsphs2}, galaxy groups~\cite{anderson15} and in the X-ray background~\cite{sekiya16}.

The  DM decay line signal in an X-ray telescope appears on top of an unrelated  astrophysical instrumental background which typically consists of continuum emission and a set of atomic lines. The signal-to-noise ratio (SNR) could be maximised with a suitable choice of the observation target. The strongest Milky Way signal typically occupies the entire field-of-view (FoV) of an X-ray telescope. Larger FoV provides higher signal statistics. The FoVs of existing telescopes operating in the 0.1-10~keV band (XMM-Newton, Chandra) are limited to a fraction of a degree. In this case a further boost of the signal could be achieved by choosing an observation direction which contains, apart from the Milky Way, also a signal from a nearby DM halo, such as e.g. a dwarf spheroidal galaxy or a galaxy cluster \cite{strategy}. To the contrary, the DM decay signal detectable by large FoV telescopes operating in the hard X-ray band above 15~keV, like e.g. ISGRI and SPI telescopes on board of INTEGRAL satellite  \cite{spi}, or of the GBM detector on board of Fermi satellite \cite{gbm} is completely dominated by the Milky Way flux.  

\nus telescope \cite{nustar} provides a large effective collection area (compared to INTEGRAL/SPI), of the order of $10^3$~cm$^2$ in the energy band above 10~keV. This is achieved with the focusing optics, as opposed to the coded mask optics of SPI. The focusing optics also provides an advantage of low background for the observations of point or mildly extended sources (compared to the coded mask optics). The large effective area an low background make \nus competitive as a DM detector which is able to provide a higher sensitivity probe of the DM decay line signal above $10$~keV, in spite of much lower energy resolution, compared to SPI. \nus data have already been used to derive constraints on the DM decay line signal from the direction of Bullet galaxy cluster \cite{dm_nustar_bullet}. The upper bound on the DM decay time stemming from this observation is comparable to the bound previously derived from the INTEGRAL/SPI \cite{spi} and Fermi/GBM data \cite{gbm}. The DM signal considered in   \cite{dm_nustar_bullet} was the signal collected through the focusing optics of the telescope. 

In what follows we show that the \nus detector perceives a much stronger DM signal (compared to that of a galaxy cluster, or of a dwarf spheroidal galaxy) in any astronomical observation. This signal originates from the Milky Way galaxy and is collected not through the focusing optics of the telescope, but rather through the aperture of the X-ray detector unit. We show that account of this signal allows to improve the sensitivity of the DM search with \nus by up to an order of magnitude in the energy range around 10~keV.  The Milky Way signal is distributed all over the sky and all the \nus pointings could, in principle, be used for the search of the DM signal. We demonstrate this by analysing two deep, several Msec long, observations of outskirts of our galaxy (COSMOS and ECDFS fields) which were previously used for the analysis of the background of \nus \cite{wik14}. We show that non-detection of the DM decay line in these observations imposes tight constraints on the mixing angle of sterile neutrino DM with masses in the range 10-30~keV. The \nus bound rules out the $\nu$MSM sterile neutrino DM with the mass higher than 20~keV.

\section{Search for the DM decay signal with \nus}
\label{sec:signal}
\subsection{Observations and data processing}

The Nuclear Spectroscopic Telescope Array (\nus) mission~\cite{nustar} was launched in 2012 and operates in energy band 3 -- 79~keV. At all energies the energy resolution of \nus is better than 10\% which makes it specifically useful for the dark matter decay line search.

The dataset used in our analysis is the extended ``blank sky fields'' dataset of~\cite{wik14}. It consists of two very deep \nus observations of COSMOS ($\sim 1$~deg$^2$ at the direction $(\ell, b)= (236,42)$) and ECDFS ($\sim 0.3$~deg$^2$ at the direction $(\ell, b)= (223, -54)$) fields. The COSMOS field contains 109 observations longer than 1~ksec each with zero issue flag taken during December 2012 -- April 2014. The ECDFS field contains 31 observations taken during September 2009 -- April 2013.

The raw data were processed with the standard pipeline processing (HEASOFT v.6.17 with  \nus subpackage v.1.5.1). Following~\cite{wik14} we have additionally applied stricter criteria for exclusion of data taken through the South Atlantic Anomaly(SAA) and a ``tentacle''-like region of higher activity near part of the SAA, producing the Level 2 data products, with the \texttt{nupipeline} tool with the flags \texttt{SAAMODE=STRICT} and
\texttt{TENTACLE=yes}. The total cleaned exposures for COSMOS and ECDFS fields are $\sim 5.0$~Msec and $\sim 2.5$~Msec ($\sim 5.5$~Msec and $\sim 2.8$~Msec without SAA removal) for the two (A+B) \nus detectors.

The high-level spectral products (spectra, response matrix and auxilary response files) were extracted from the central $6'$ radius region with \texttt{nuproducts} routine with \texttt{extended=yes} flag, most appropriate for extended sources\footnote{See e.g. \url{http://heasarc.gsfc.nasa.gov/docs/nustar/analysis/nustar_swguide.pdf}} and \texttt{bkgextract=no} flag, since we  had an aim to model of the instrumental and astrophysical background.
For the further analysis we have added considered observations, producing one spectrum per camera per observational field with the \texttt{addspec} routine resulting in 4 spectra, referring hereafter as ``all dataset''. To reduce possible contamination of the data by the Sun at low ($\lesssim 10$~keV) energies we have  also considered a subset of these data taken while \nus was shadowed by the Earth (``no Sun'' dataset). The ``no Sun'' dataset event files were obtained from ``all'' dataset files by filtering with \texttt{nuscreen} routine for \texttt{ELV>5} and \texttt{SUNSHINE==0} expressions and invoked with the \texttt{cleancols=no} flag in order to prevent removing of the auxiliary data from event files. Further processing of ``no Sun'' dataset was performed in a way similar to ``all'' dataset with \texttt{nupipeline} and \texttt{nuproducts} routines.

For each dataset we have performed a search for narrow gaussian lines in energy range 3-70~keV above the model consisting of instrumental and astrophysical background components. 

The instrumental background model consists of a power law and 22 gaussian lines in 3 -- 70~keV band with the positions adopted from Ref.~\cite{wik14}. For each of the fitted spectra we have additionally adjusted parameters of the background model allowing the line centroids and dispersions to vary within the line width given in~\cite{wik14}. During the fitting the instrumental background model was not convolved with auxiliary response.
 
\begin{figure*}
\includegraphics[width=0.35\textwidth, angle=-90, origin=br]{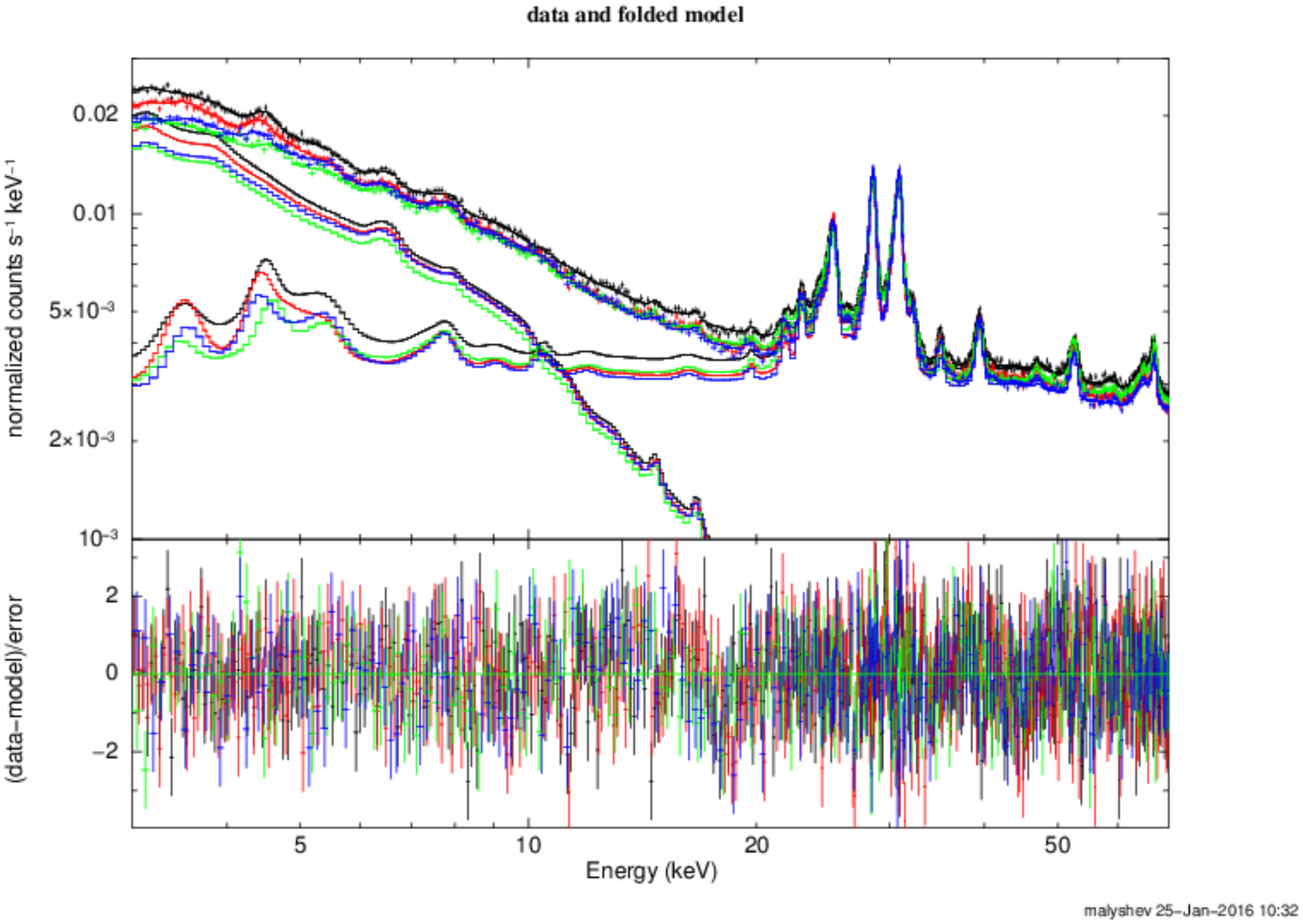}
\includegraphics[width=0.45\textwidth]{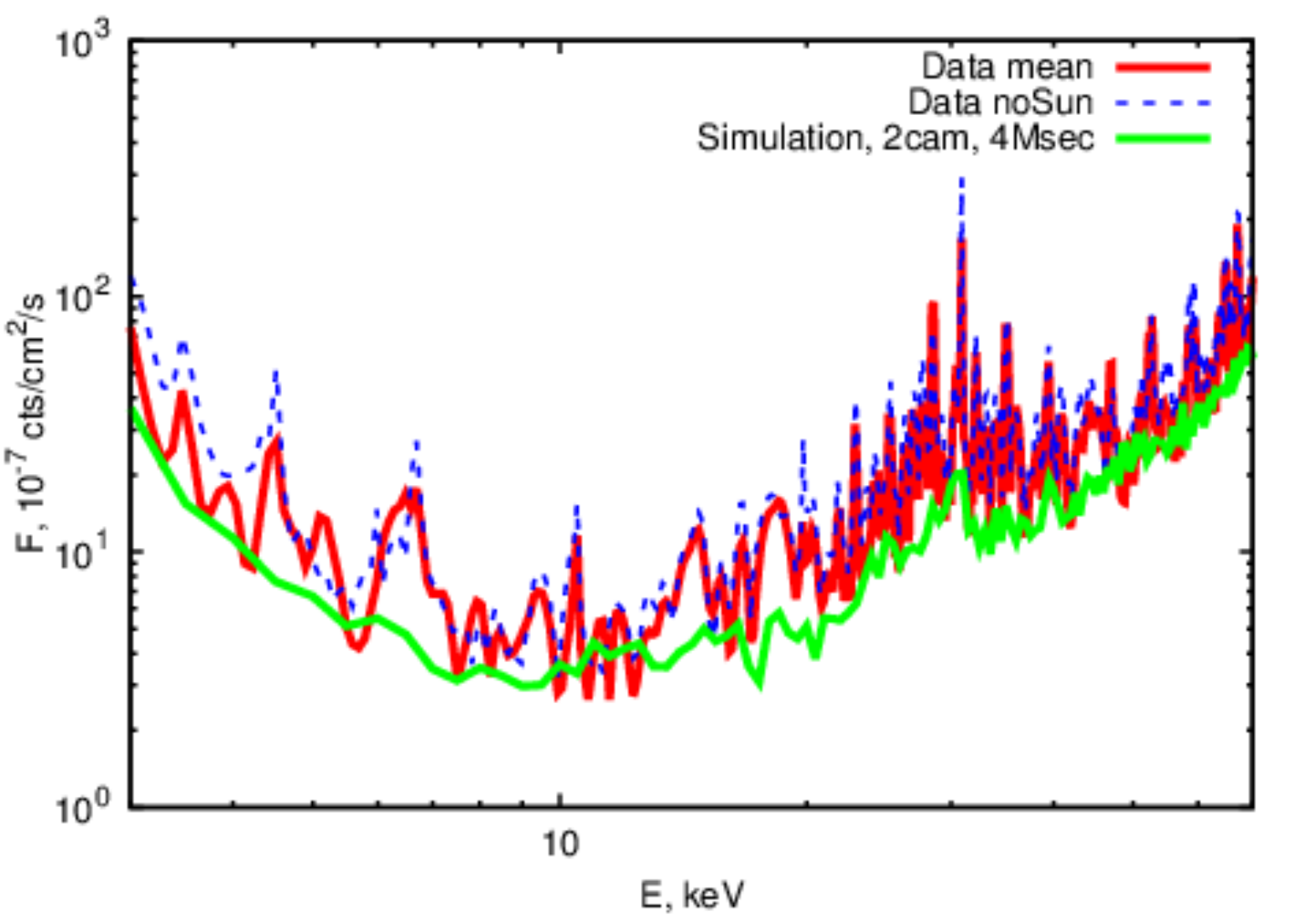}
\caption{Left: COSMOS and ECDFS fields spectra in 3-70~keV energy band with the best-fit considered background model. Right: $3\sigma$ upper limits on a narrow gaussian line flux added on top of the background model. The results for ``all'' and ``no Sun'' datasets are shown with solid red and dashed blue lines correspondingly. The solid green line shows the upper limits estimated from simulations~\cite{we_athena}, see text for the details.}
\label{fig:spec_flux}
\end{figure*}

The astrophysical background model includes solar (\texttt{apec}) and diffuse x-ray background (broken power law) components with free normalization/slope/temperature parameters. The high energy slope of the broken power law component with fixed to best-fit value 1.68 observed by INTEGRAL/ISGRI at energies $\gtrsim 20$~keV~\cite{integral_cxb,bat_cxb,isgri_cxb}. The best-fit low energy slope $\Gamma_1=1.42 \pm 0.25$ is in agreement with CXB slope seen by XMM-Newton and Chandra~\cite{deluca04,kev_cxb}. We locate the break in the spectrum to be present at energy $E_{br}=14.0 \pm 1.2$~keV.
``All'' dataset spectra together with the best-fit model are shown in Fig~\ref{fig:spec_flux}, left.

$3\sigma$ upper limit on the flux of added line as a function of the line centroid for ``all'' and ``no Sun'' datasets is shown in Fig.~\ref{fig:spec_flux}, right. The results for ``all'' and ``no Sun'' datasets are shown with solid red and dashed blue line correspondingly. The negative best-fit values of the line flux were treated as the systematics and added (linearly) to the shown upper limits. For comparison, we also show in the same figure an estimate of sensitivity of \nus calculated for a comparably long exposure in the direction of a dwarf spheroidal galaxy~\cite{we_athena}.

\subsection{Detected lines}

The list of lines detected with the significances $>3\sigma$ in the energy range 3-20~keV together with lines empirically included into the instrumental background model  of Ref. \cite{wik14} is given in Table~\ref{tab:lines}. We do not list the lines detected above 25~keV because there is little doubt that these lines are of instrumental rather than astrophysical origin. In this energy range SPI instrument of INTEGRAL telescope provides superior sensitivity to the astrophysical line flux \cite{spi}, compared to \nus.  Between 20 and 25 keV the \nus background is dominated by overlapping multiple instrumental lines as described in the Ref. \cite{nustar}. This makes the search of the dark matter line in this energy range difficult. 

For all the  lines present in the Table~\ref{tab:lines} we list line energy (keV) and its detection significance (standard deviations, $\sigma$), best-fit width of the line (fixed to 0. for ``sky'' background lines), fluxes for ``all'' and ``no Sun'' datasets and the information on possible origin of the line. The assignment of the majority of the lines to ``instrumental'' or ``sky'' background is not obvious. We tentatively list the line as ``instrumental''  if it is present in the instrumental background model of~\cite{wik14} or has best-fit width is largely inconsistent with zero. Equally uncertain is elemental identification of most of the lines. The identifications suggested in the right column of Table ~\ref{tab:lines} should be considered as "suggestive" rather than certain identification. The detected lines could also be "line-like features" which occur at energies where modelling of the continuum components of the background is not precise enough. Below we provide additional comment on some of the detected lines and line like features.

\subsubsection{3.5~keV  line} 

This line is present in the instrumental background model of~\cite{wik14} and is attributed to reflection of the sunlight on the telescope structure. We have attempted a verification of hypothesis of the solar origin of the line by
considering the ``no Sun'' dataset. In this dataset we find that the flux in 3.51~keV line remains constant within statistical uncertainty. This makes the solar origin of the line unlikely. 

The feature appears as a narrow line (the width is consistent with 0 at $<2\sigma$ c.l.). This line energy is at the edge of \nus energy range, where large uncertainties of response functions are potentially present. Addition of the line might be favoured by the model fit because it allows to "compensate" the imperfection of the response functions. 

A line with similar energy was observed recently in the stacked spectra of galaxy clusters, of M31 galaxy and in the Galactic Center~\cite{line35_1,line35_2,line35_gc,iakubovskyi15}  (see, however~\cite{horiuchi:14,riemer:14,potassium,dsphs,anderson15,sekiya16,hofmann16} and~\cite{dsphs2,bulbul16}), which makes this line of particular interest for a deeper study. Assuming that the line is of astrophysical origin, one could estimate  the line flux, $7.7 \pm 1.3$~cts/s/cm$^2$. This flux is somewhat higher than typical estimates of the Milky Way flux from the directions of COSMOS and ECDFS fields,  but lies within the uncertainty range of the Dark Matter decay line flux estimates which include the uncertainty of the knowledge of the DM column density in these directions and the uncertainty of the parameters of the dark matter particles derived from the analysis of the galaxy clusters and M31~\cite{line35_1,line35_2}.

\subsubsection{Sun modulated lines} 

Considering ``all'' and ``no Sun'' datasets we have found strong flux variation of the  lines associated to the presence of the Sun illuminating the telescope. These are the lines at  4.46~keV, 6.32~keV and 7.96~keV.  The 4.46~keV line is mentioned as the Sun-excited line in the Ref. \cite{wik14} and  could be identified with Ti K$\alpha$ line because Titanium is present in the spacecraft material\footnote{we thank Dr. F.~Gastaldello for pointing this to us.}. We make a conjecture that similarly to the line at 4.46~keV, the lines at 6.32 and 7.96~keV  are also produced via excitation of the spacecraft material by the sunlight.

\subsubsection{"Ghost" lines} 

An additional possibility for appearance of a line-like feature stems form a complicated shape of the \nus energy Redistribution Matrix (RMF) which has a strong low-energy "wing". This wing might appear in a form of a "ghost" line in the spectrum if a much stronger line is present at higher energy and modelling of the RMF suffers from limited precision.   In Table \ref{tab:lines} we have marked the energies of possible "ghost" lines for all the detected strong  ($>40\sigma$ significance) lines found in the spectrum. The lines appearing at the positions of the possible "ghosts" are marked as such in Table \ref{tab:lines} together with the energy of the strong "parent" line of the ghost.   

\subsubsection{Instrumental lines} 

The instrumental lines are produced by the material of the telescope and spacecraft. They could be excited by interactions of cosmic rays with the spacecraft and also by illumination of the spacecraft (e.g. with the Sun). The most straightforward identification could be done for the K$\alpha$ lines which are the lines for the excitations by cosmic rays. Table \ref{tab:lines} shows which of the detected lines could be the K$\alpha$ lines of certain elements.  The line candidate identification is based on Ref.~\cite{bearden67} and the atomic lines database\footnote{See e.g. \tt http://www.kayelaby.npl.co.uk/atomic\_and\_\\ nuclear\_physics/4\_2/4\_2\_1.html}.



\begin{table*}
\begin{tabular}{|c|c|c|c|c|c|c|c|}
\hline
Line energy, & Significance & Width, & $F$, & $F_{noSun}$,& Sun? & Ghost?& Comments \\
keV & $\sigma$ & keV & $10^{-6}$~cts/cm$^2$/s & $10^{-6}$~cts/cm$^2$/s & && \\
\hline
3.51*  $\pm$ 0.02 & 11.1 & 0.08 $\pm$ 0.05 & 7.7 $\pm$ 1.3  & 10 $\pm$ 2.5 & &&lower edge of \\
&&&&&&&sensitivity band\\
4.46* $\pm$ 0.05 & 15.7 & 0.12 $\pm$ 0.03 & 5.9 $\pm$ 0.5 & 3.7 $\pm$ 0.5 &Y&&  Ti $K\alpha$  \\
4.7* $\pm$ 0.1 & 9.8 & 0.6 $\pm$ 0.1 & 8.9 $\pm$ 1.8 & 8.2 $\pm$ 1.9 & &&\\
6.32 $\pm$ 0.08 & 6.7 & 0. & 1.2 $\pm$ 0.2 & 0.66 $\pm$ 0.23 & Y &&Fe $K\alpha$ ? \\
7.96 $\pm$ 0.06 & 4.0 & 0. & 0.5 $\pm$ 0.1 & 0.23 $\pm$ 0.18  & Y&&Cu $K\alpha$ ?\\
10.44* $\pm$ 0.05 & 8.9 & 0.2 $\pm$ 0.05 & 1.4 $\pm$ 0.2& 1.7 $\pm$ 0.3 & && W L-edge residuals \cite{calib} \\
14.2 $\pm$ 0.1 & 3.3 & 0. & 0.51 $\pm$ 0.18 & 0.6 $\pm$ 0.2  & &&Sr $K\alpha$?\\
14.75 $\pm$ 0.05 & 5.9 & 0. & 0.9 $\pm$ 0.2 & 1.0 $\pm$ 0.2 &  & Y? &23~keV ghost?\\
15.7 $\pm$ 0.1 & 3.7  & 0. &0.57 $\pm$ 0.16&0.6 $\pm$ 0.2 &  &Y? &24.5~keV ghost,  Zr $K\alpha$?\\
16.7 $\pm$ 0.1 & 5.5 & 0. & 0.9 $\pm$ 0.2 &1.2 $\pm$ 0.2 & &Y?&25.3~keV ghost, Nb $K\alpha$? \\
19.66* $\pm$ 0.06 & 9.3 & 0.06 $\pm$ 0.14 & 1.3 $\pm$ 0.3 & 1.3 $\pm$ 0.3 & &Y?& 28.5~keV ghost? \\
\hline
\end{tabular}
\caption{The lines detected at $\gtrsim 3\sigma$ level above the background model, for ``all'' and ``no Sun'' datasets in 3-20~keV range. Lines presented in the instrumental background model are marked with ``*''. The parameters of all lines (including instrumental) were derived from the best-fit in ``all'' dataset, assuming equal parameters in all spectra and sky origin. The line significance is determined from the difference of the statistics of the model with and without the corresponding line. ``Comments'' column lists the elements with fluorescence emission line or an absorption edge within the specified uncertainty for the line position. ``Ghost''=Y flag indicates the possible contamination originating from the non-linearity of \nus response and the presence of strong narrow line at higher energy.}
\label{tab:lines}
\end{table*}

\subsection{Constraints on parameters of the dark matter particles}

The focused X-ray signal from observed sources enters the detector of \nus telescope through an aperture opening which also lets non-focused X-ray background in \cite{nustar}.  For the standard point-source analysis this ``aperture'' background is estimated and subtracted.

The aperture background is composed of diffuse X-ray emission from extragalactic and Galactic sources within a sky region of the area   $\sim 37.2$~deg$^2$ around the telescope axis (compare with $13'\times 13'$ focused FoV). This diffuse X-ray emission potentially includes also the flux of the dark matter decay line originating from the Milky Way dark matter halo. Subtraction of the aperture background would decrease the dark matter signal. To avoid this, we model the  dark matter signal coming to the detector from both the focusing optics and the detector aperture window.

\begin{figure}
\includegraphics[width=0.45\textwidth]{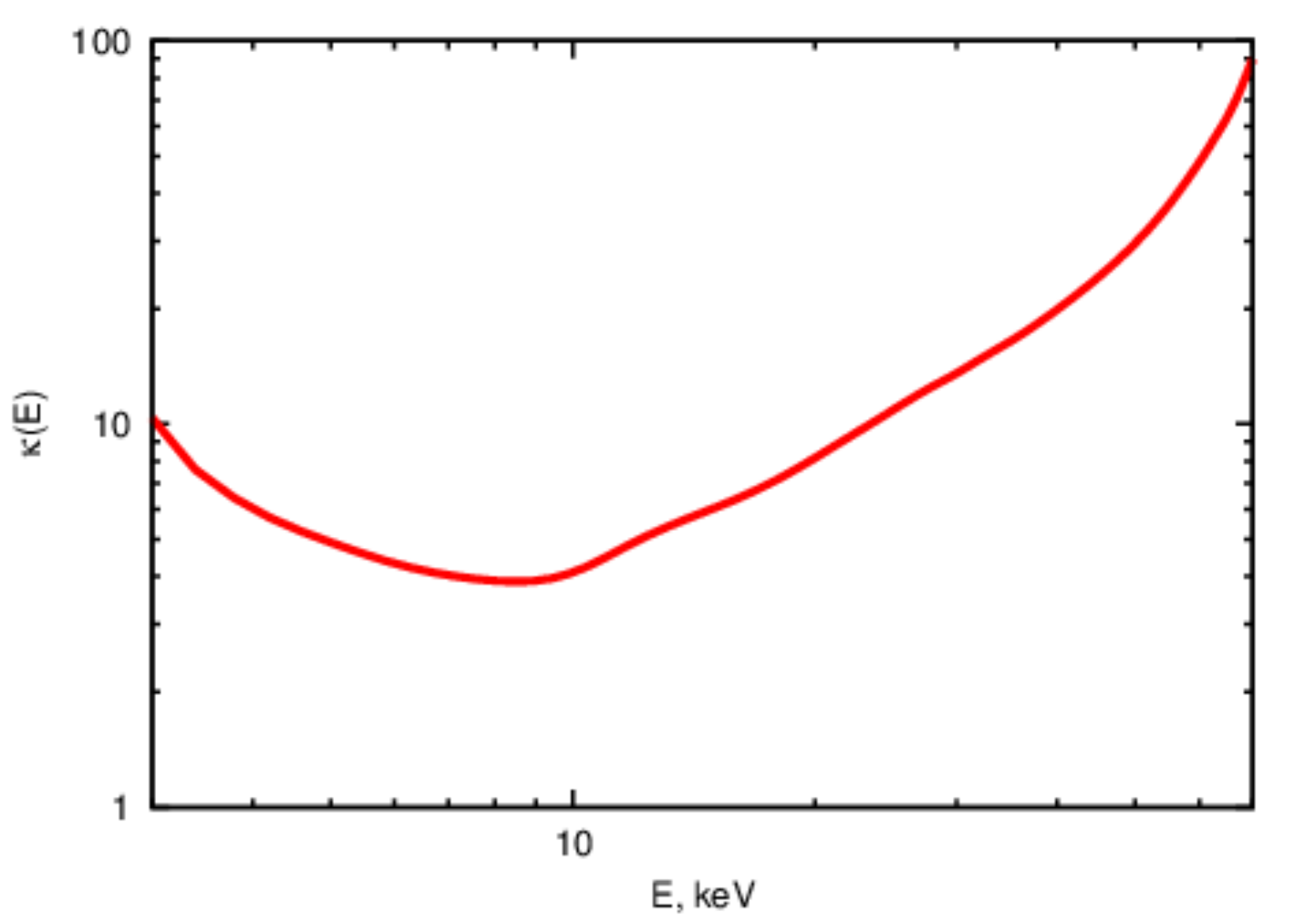}
\caption{The ratio of the aperture and the focused parts of the dark matter signal as a function of energy.}
\label{fig:kappa}
\end{figure}

The count rate in the dark matter decay line is given by the sum of the  aperture and focused components:
\begin{equation}
{\cal R}_{DM} = F_{ap}A_{ap} + F_{foc}A_{foc} = F_{obs}A_{foc}
\end{equation}
where $F_{ap}, F_{foc}$ and $F_{obs}$ are the  aperture, focused and apparent DM decay fluxes, $A_{ap},A_{foc}$ are the aperture and focused signal effective areas. Inclusion of the aperture component of the signal boosts the  apparent dark matter signal by a factor 
\begin{equation}
\kappa(E) \equiv F_{obs}/F_{foc} = 1+\frac{\Omega_{ap}A_{ap}(E)}{\Omega_{foc}A_{foc}(E)}
\end{equation}
compared to the signal coming only through the focusing optics.  The quantity $\kappa(E)$ is plotted as a function of energy in Fig.~\ref{fig:kappa}. To produce this plot, we have extracted the effective area from auxiliary response file for A camera for COSMOS field observations and from auxiliary file used for the modelling the aperture background within \texttt{nuskybgd}\footnote{See e.g. https://github.com/NuSTAR/nustar-idl/tree/master/nuskybgd} package, provided by \nus collaboration. As can be seen from Fig.~\ref{fig:kappa}, account of the unfocused aperture X-ray signal increases the expected DM signal in \nus observations by factor $\sim 4-100$.

The COSMOS and ECDFS fields are located at approximately the same Galactic longitude and at opposite Galactic latitudes.  This "symmetric" location leads to approximately equal estimates of the dark matter column densities $S$ in the two fields.  The estimations of Ref. \cite{klypin01} give preferred (minimal) column densities values of $7.2\cdot 10^7 M_{\bigodot}/pc^2$ ($3.4\cdot 10^7 M_{\bigodot}/pc^2$) for the two fields. More recent works~\cite{mcmilan11,read14,pato15}, suggest a somewhat  higher estimates $10 - 12 \cdot 10^7 M_{\bigodot}/pc^2$. In what follows we adopt ``preferred'' value from~\cite{klypin01} as a conservative estimation and notice that there is a systematic intrinsic uncertainty of about a factor of $\simeq 2$ of this estimate.

The dark matter column density determines the decay line flux from dark matter sterile neutrino~\cite{strategy,dsphs}
\begin{eqnarray}
F_{MW} = 10^{-7}\kappa\left[\frac{\theta}{6'}\right]^2\left[\frac{S}{10^{22}\mbox{GeV/cm}^2}\right] \\ \nonumber
\left[\frac{\sin^2(2\theta_{DM})}{5\times 10^{-11}}\right]\left[\frac{m_{DM}}{7\mbox{ keV}}\right]^4~\mbox{ph/cm}^2/\mbox{s}
\label{eq:F_DM}
\end{eqnarray}  
Here $m_{DM}$ is the mass of the dark matter particle and $\theta_{DM}$ is the mixing angle which determines the strength of the coupling of the dark matter to conventional matter and  $\theta=6'$ is the radius of the region in the focused field-of-view selected for the analysis. The factor $\kappa$ takes into account the additional signal entering through the detector aperture window, as discussed above.

\begin{figure}
\includegraphics[width=\linewidth]{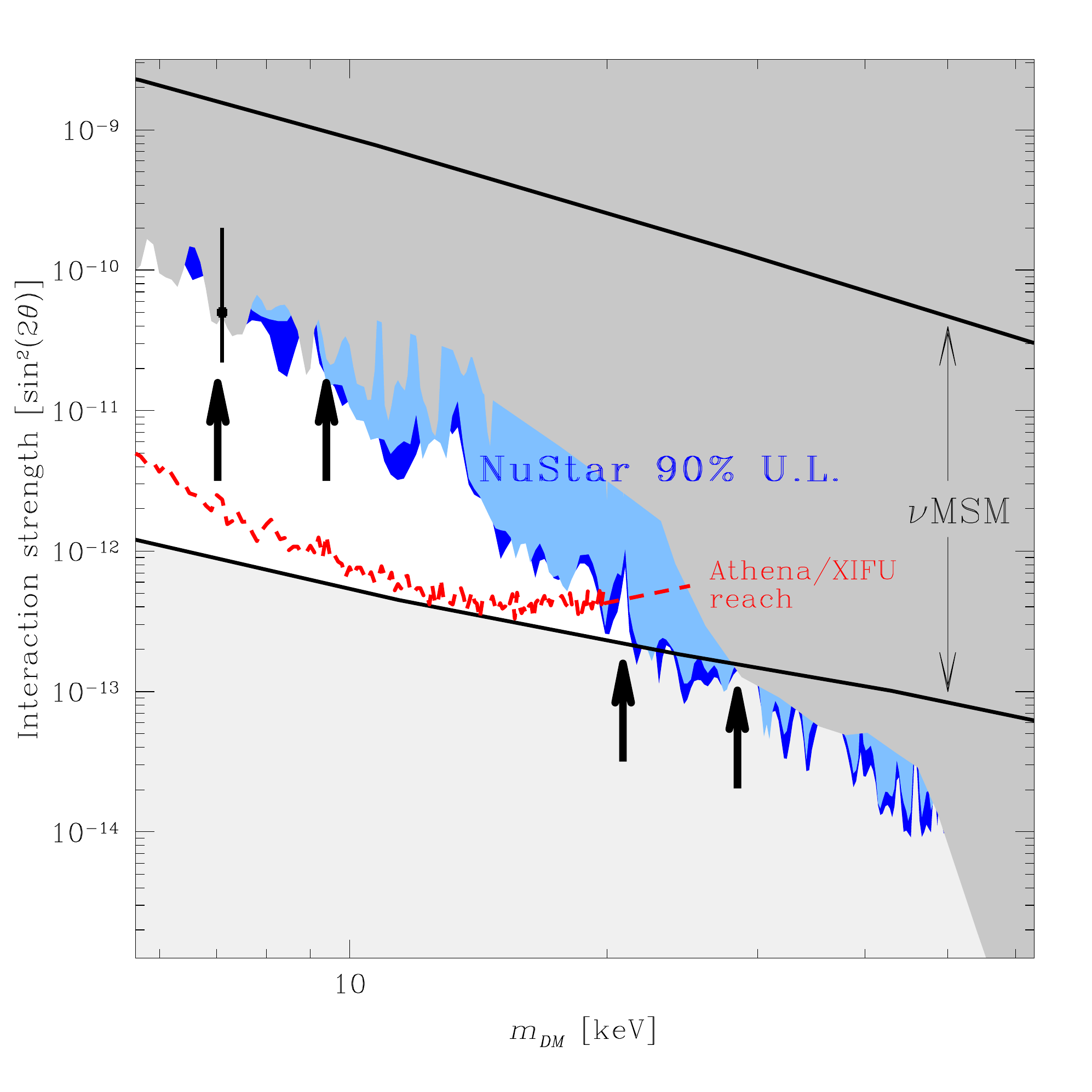}
\caption{Constraints on sterile DM mass $m_{DM}$ and mixing angle $\sin^2(2\theta_{DM})$ from \nus observations of COSMOS and ECDFS fields. Light and dark blue shadings show the results from ``all'' and ``no Sun'' datasets. Darker grey shows previously derived constraints from X-ray observations. Black lines delimit the allowed range of parameters of $\nu$MSM model. Dashed red curve shows the sensitivity reach of Athena/XIFU observations of dwarf spheroidal galaxies. Data point marks the estimate of the dark matter sterile neutrino mixing angle for the dark matter interpretation of  the 3.5~keV line in galaxy clusters and in M31. Upward arrows mark the masses of dark matter particles which would produce the $\gtrsim 3\sigma$ lines found in the \nus spectra of COSMOS and ECDFS fields. }
\label{fig:constraints}
\end{figure}

Comparing the expected dark matter decay flux with the upper limits derived from the analysis of the COSMOS and ECDFS data, shown in Fig.~\ref{eq:F_DM}, one could derive constraints on the dark matter sterile neutrino mixing angle $\theta_{DM}$. These constraints are shown in Fig.~\ref{fig:constraints}. 

Non-detection of the line signal at all energies except for the energies listed in Table \ref{tab:lines} improves previously derived constraints in the dark matter particle mass range between 7 and 40~keV. The improvements of limits on $\theta_{DM}$ are particularly strong, by an of magnitude, in the energy range between 10 and 30~keV. In this energy range, the tightest constraints were previously imposed by the non-detection of the dark matter decay line in a deep observation of Bullet galaxy cluster with \nus \cite{dm_nustar_bullet}. 

The improvement of sensitivity of the dark matter line search compared to this previous analysis could be readily explained.  Presence of a large dark matter overdensity in the  field-of-view of the telescope (a massive galaxy cluster)  in the dataset analysed in Ref. \cite{dm_nustar_bullet} boosts the expected line flux (for a given $\theta_{DM}$) and makes the detection of the signal easier. However, as it is shown in Ref. \cite{strategy}, the dark matter decay signal from the Milky Way halo is nearly as strong as the signal from typical distant massive dark matter halos, like galaxy clusters. Thus, the background X-ray flux visible in the telescope typically carries the signal of comparable strength. In the particular case of \nus , an additional contribution to the Milky Way halo signal is given by the non-focused X-ray flux falling on the detector through the aperture opening. We have shown that this addition boosts the signal by one-to-two orders of magnitude, depending on photon energy.  In addition, the exposure of the COSMOS and ECDFS fields is an order of magnitude longer than that of the Bullet cluster. The combined effects of the boost of the decay signal by the aperture component and longer exposure result in better sensitivity for the dark matter decay line search.

\section{Discussion}

Our search of narrow lines in the \nus background spectrum has resulted in detection of a set of lines listed in Table \ref{tab:lines}. Some of these lines are clearly of instrumental origin, such as e.g. the lines at 4.5, 6.4 and 8 keV excited by illumination of the spacecraft by the Sun. The origin of other   lines is less clear. 

The lines in the energy range 14-20~keV are likely to be residuals from improper modelling of the \nus RMF. This RMF has a low-energy "wing" which could result in "ghost" lines at the low-energy wings of the strong instrumental lines. The \nus\ background in the 20-30~keV band is dominated by a blend of strong instrumental lines \cite{nustar}. Our testing of the hypothesis of "ghost" origin of the lines has resulted in suggestive association of the lines above 14~keV with "ghosts" of the stronger instrumental lines at higher energies.

The line at 10.4~keV is most probably also of instrumental origin. The effective area of \nus experience a large jump at the L-edge of tungsten. The fit of the background model might favour inclusion of an additional line at the energy of the edge to compensate for the defects of the modelling of the edge shape in the effective area. 

Two lines: at 3.5 and 4.7~keV have less obvious origin. Although the 3.5~keV is mentioned as being of solar origin in the Ref. \cite{wik14}, we do not find the expected variation of the line flux between the ``sun-illuminated'' and ``no-sun'' parts of the data set. The most probable origin of this line is the imperfection of the modelling of the energy dependence of the \nus effective area close to the low energy threshold of the instrument (nominally at 3~keV). At the same time, we could not exclude the possibility of the dark matter origin of the line. We also have not found a plausible explanation for the 4.7~keV line.\footnote{Note however, that $Cs L\beta$  line can be produced at similar energy and stronger Cs $K\alpha$, $K\beta$ lines are present in instrumental model at $\gtrsim 30$~keV band which may indicate the possible instrumental origin of 4.7~keV line.}

The alternative hypotheses of instrumental and astrophysical (dark matter decay) origin of the detected lines could be tested using the strategy described in the Ref. \cite{spi}. The intensity of the line originating from the dark matter halo of the Milky Way is expected to vary across the sky exhibiting an excess toward the inner Galaxy and deficit of the flux from the outer Galaxy. To the contrary, the instrumental line is not expected to show clear large scale variability pattern on the sky (spurious variations could still be induced by the changes of observational conditions during exposures of different parts of the sky). A test of the dark matter origin of the detected lines could therefore be done with a set of additional "deep field" exposures at different off-Galactic-centre directions.

An alternative approach for the testing of the dark matter origin of the lines is to obtain deep exposures of selected dwarf spheroidal galaxies (e.g. Draco). This should provide a factor-of-two boost in the expected dark matter line signal for the real dark matter decay line (as derived in Ref. \cite{we_athena}) and no variation of the signal for the instrumental line. 

The improvement of sensitivity of the search of the dark matter decay line provided by the \nus exposures of COSMOS and ECDFS fields is important in the context of testing the reference $\nu$MSM model of sterile neutrino dark matter. Fig. \ref{fig:constraints} shows that the \nus data limit the mass of the dark matter neutrino to be below 20~keV, otherwise production of sufficient amount of the dark matter in the Early Universe would require too high lepton asymmetry. Exclusion of the 20-30~keV mass range closes a sensitivity gap of future X-ray telescope Athena, which will have an X-ray spectrometer XIFU able to detect the dark matter decay line in the energy range below 10~keV (neutrino mass range below 20~keV)  \cite{we_athena}.

\label{sec:conclusions}

\section*{Acknowledgements}
The authors would like to thank Dr. Fabio Gastaldello for useful comments on the text.

\def\aj{AJ}%
\def\actaa{Acta Astron.}%
\def\araa{ARA\&A}%
\def\apj{ApJ}%
\def\apjl{ApJ}%
\def\apjs{ApJS}%
\def\ao{Appl.~Opt.}%
\def\apss{Ap\&SS}%
\def\aap{A\&A}%
\def\aapr{A\&A~Rev.}%
\def\aaps{A\&AS}%
\def\azh{AZh}%
\def\baas{BAAS}%
\def\bac{Bull. astr. Inst. Czechosl.}%
\def\caa{Chinese Astron. Astrophys.}%
\def\cjaa{Chinese J. Astron. Astrophys.}%
\def\icarus{Icarus}%
\def\jcap{J. Cosmology Astropart. Phys.}%
\def\jrasc{JRASC}%
\def\mnras{MNRAS}%
\def\memras{MmRAS}%
\def\na{New A}%
\def\nar{New A Rev.}%
\def\pasa{PASA}%
\def\pra{Phys.~Rev.~A}%
\def\prb{Phys.~Rev.~B}%
\def\prc{Phys.~Rev.~C}%
\def\prd{Phys.~Rev.~D}%
\def\pre{Phys.~Rev.~E}%
\def\prl{Phys.~Rev.~Lett.}%
\def\pasp{PASP}%
\def\pasj{PASJ}%
\def\qjras{QJRAS}%
\def\rmxaa{Rev. Mexicana Astron. Astrofis.}%
\def\skytel{S\&T}%
\def\solphys{Sol.~Phys.}%
\def\sovast{Soviet~Ast.}%
\def\ssr{Space~Sci.~Rev.}%
\def\zap{ZAp}%
\def\nat{Nature}%
\def\iaucirc{IAU~Circ.}%
\def\aplett{Astrophys.~Lett.}%
\def\apspr{Astrophys.~Space~Phys.~Res.}%
\def\bain{Bull.~Astron.~Inst.~Netherlands}%
\def\fcp{Fund.~Cosmic~Phys.}%
\def\gca{Geochim.~Cosmochim.~Acta}%
\def\grl{Geophys.~Res.~Lett.}%
\def\jcp{J.~Chem.~Phys.}%
\def\jgr{J.~Geophys.~Res.}%
\def\jqsrt{J.~Quant.~Spec.~Radiat.~Transf.}%
\def\memsai{Mem.~Soc.~Astron.~Italiana}%
\def\nphysa{Nucl.~Phys.~A}%
\def\physrep{Phys.~Rep.}%
\def\physscr{Phys.~Scr}%
\def\planss{Planet.~Space~Sci.}%
\def\procspie{Proc.~SPIE}%
\let\astap=\aap
\let\apjlett=\apjl
\let\apjsupp=\apjs
\let\applopt=\ao

\end{document}